\documentclass[%draft
  ]
  {aipproc}

\layoutstyle{6x9}

\begin{document}

\title{Varying ``constants'' in cosmology and astrophysics and\ldots}


\author{Thomas Dent}{
  address={Theoretical Physics, University of Heidelberg, Philosophenweg 16, 
69120 Germany}
}

% \author{<author2>}{
%  address={<common address for author2 and author3>}
% }

% \author{<author3>}{
%  address={<common address for author2 and author3>}
%  ,altaddress={<author1 address>} % additional visiting address
% }

\begin{abstract}
We review astrophysical, cosmological and terrestrial evidence for and against 
the constancy of fundamental parameters of particle physics, and discuss 
theoretical issues of unification and scalar-mediated forces, finding that the
current rate of variation is bounded by limits on violations of the weak 
equivalence principle. 
%We also comment in verse on the general situation in particle physics. 
\end{abstract}

\maketitle

%%%%%%%%%%%%%%%%%%%%%%%%%%%%%%%%%%%%%%%%%%%%
%% MAINMATTER
%%%%%%%%%%%%%%%%%%%%%%%%%%%%%%%%%%%%%%%%%%%%

\paragraph{Introduction}
The constancy of the parameters of particle physics \cite{Uzan}, such as the fine 
structure expansion parameter $\alpha$ of electromagnetism, is an assumption 
that should be tested, since new physics has often arisen from the breakdown 
of assumptions.
%, such as the passive nature of spacetime or the inertness of the vacuum.

At present not all tests are unambiguously consistent with constancy, and 
there is a theoretical framework, involving extending particle physics models 
with one or more cosmologically varying scalar fields \cite{Wetterich:2002ic}, which can lead to 
interesting constraints and predictions. Experiments are continuously being 
improved and updated and we can expect many puzzles to be resolved within a 
few years. 

This talk summarizes the techniques and results in cosmology, astrophysics, 
nuclear physics and atomic physics in investigating the alleged constancy of 
physical parameters, and outlines a current research project on primordial
nucleosynthesis (BBN). We also focus on two theoretical questions: first, 
whether we can relate independently measurable parameters such as $\alpha$ 
and the proton-electron mass ratio $\mu$ in unified theories; second, what
long-range forces may result from the scalar fields which are posited to
be the source of nonzero variation?

\paragraph{Relation to postulates of General Relativity}
The variation of (locally) measurable physical parameters over spacetime 
violates Local Position Invariance which is part of the Einstein Equivalence 
Principle. Still, ``varying constants'' may be studied within a generally 
covariant theory where scalar degrees of freedom are excited. As we will see, 
the universality of free fall, another essential postulate in GR, may also be 
violated due to the scalar coupling to matter. 

\paragraph{Atomic and molecular transitions}
The most exact and direct measurements of fundamental parameters come from
electromagnetic transitions in atoms and molecules, measured by various 
optical techniques. Since all such transitions are proportional to the 
Rydberg constant one must measure two or more to constrain the value of a 
dimensionless parameter. Different transitions have different functional 
dependence on $\alpha$ and $\mu\equiv m_p/m_e$ (and on magnetic moments 
%of particles and nuclei - see for example 
\cite{Karshenboim}).

\paragraph{Atomic clocks} By measuring two or more different atomic 
transitions over several years (for example the SI frequency standard, a 
hyperfine transition of ${^{133}}$Cs) one can limit the present rate of change 
of various fundamental parameters. Limits on $\dot{\alpha}/\alpha$ of $3\times
10^{-15}$ per year have been achieved \cite{Fischer} using atomic hydrogen, 
mercury and caesium, and $2\times 10^{-15}\,$y$^{-1}$ with transitions of
a single Yb$^+$ ion \cite{Peik04}.

\paragraph{Astrophysical spectra} Quasar absorption spectra offer a means to
probe the values of fundamental parameters over cosmological time, though with
much less absolute accuracy than atomic clocks. One looks for absorption 
lines which have a high sensitivity to $\alpha$ (or some other parameter) and 
are optically distinct and not saturated. Since the
redshift of each system is a priori unknown, at least two 
% and preferably more
transitions must be measured. Fitting to the velocity profiles of absorption 
systems introduces an uncertainty whose size is debatable in each individual 
system but should average to zero over many systems (as should differential
velocities between different species).
A nonzero fractional variation of $\alpha$ is claimed \cite{pro} at the level 
$(-0.57 \pm 0.11) \times 10^{-5}$ (average over 143 systems with $0.2<z<4.2$); 
this is contradicted by null results with quoted accuracy down to 
$0.15\times 10^{-5}$ derived from a relatively small number of systems 
\cite{novary}. 

Recently the proton-electron mass ratio was measured to have varied
significantly from the current value: the claimed fractional variation 
\cite{mu} is
$(2.4 \pm 0.6)\times 10^{-5}$, from two molecular hydrogen systems at redshift 
around 3. Other dimensionless physical parameters may also be probed, and
null results have been obtained for the fractional variations of the products 
$\alpha^2 g_p$ and $\alpha^2 g_p/\mu$ at an accuracy of $10^{-5}$ 
\cite{otherastro}. 

\paragraph{Nuclear physics} Nuclear phenomena are more complicated than atomic 
or molecular transitions since they involve the strong nuclear force, which is 
now understood as the residual effects of QCD acting between particles with 
confined colour charges. The QCD confinement scale $\Lambda_c$ can be taken as 
a fundamental energy scale or unit for hadronic and nuclear processes; then 
the relevant parameters are $\alpha$ and the Fermi constant $G_F$, plus the 
light quark masses which affect the masses and interactions of both hadrons 
and mesons. Most nuclear reactions involve more than one type of interaction, 
thus they depend nontrivially on more than one parameter leading to possible 
degeneracies. 
%In addition, although dependence on $\alpha$ may be estimated with some confidence, 

Nuclear physics effects at distant epochs can only be probed indirectly, via 
astronomical measurements of relative isotopic abundances, or measurements of 
asteroids or rock samples on Earth.
%where there is some independent estimate of the time scale over which reactions took place. 

\paragraph{Oklo} Isotopic ratios of many elements in the Oklo uranium mine in 
Gabon differ strikingly from values obtained elsewhere on Earth, indicating 
that extensive nuclear reactions occurred there at some past epoch. The higher
fraction of ${^{235}}$U in the past, combined with an unusual rock formation 
and water moderation of neutrons allowed a fission chain reaction to occur. 
By comparing isotopic ratios of different elements one can deduce whether their
cross-sections for neutron capture $\sigma_n$ (averaged over an estimated
neutron energy distribution) had the same ratio at the time of the reactor 
operation (1.8 billion years ago, $z\simeq 1.3$) as today. Then if the 
dependence of $\sigma_n$ on ({\em e.g.}) $\alpha$ is known for different 
isotopes, a bound on $\alpha$ can be deduced up to possible degeneracies. The 
strongest bound arises from the ${^{149}}$Sm$/{^{147}}$Sm ratio: ${^{149}}$Sm 
has a sharp neutron capture resonance whose $\alpha$ dependence is enhanced by 
accidental cancellation of nuclear {\em vs.} electromagnetic energy; the 
fractional variation of $\alpha$ is thus limited below $10^{-7}$ 
\cite{Damour:1996zw}. 
%most recent analysis the fractional change in is limited to $6\times 10^{-8}$ \cite{Petrov}.

Other bounds can be obtained from considering decays in meteorites believed
to have formed around the same time as the Solar System \cite{Olive:2003sq}, although these are
also subject to degeneracy and can only test the averaged values of parameters 
over billions of years. Nuclear data can also be interpreted as bounding the
variation of quark masses, but the dependence of nuclear forces on quark masses
is still subject to much theoretical uncertainty, requiring further efforts in 
lattice and effective field theory.

\paragraph{Primordial nucleosynthesis (BBN)} The isotopic composition of matter
in the early Universe is a witness to nuclear reactions that proceeded in the 
hot plasma soon after the Big Bang. Starting with protons and neutrons in 
equilibrium, models based on laboratory measurements of cross-sections are 
used to track the progress of reactions and predict the resulting abundances 
of light elements.
% when these reactions freeze out. 
These can be compared with 
astronomical observations which attempt to measure nuclear abundances 
%of light elements 
in stars or gas and extrapolate back to a point where the effects of 
astrophysical processing were negligible. The clearest test is deuterium 
%(usually written as D$/$H) 
which is only destroyed in astrophysical processes, hence any measurement of 
D$/$H is a lower bound on the primordial value. Other isotopes considered are 
${^4}$He, which accurately reflects the neutron-proton ratio at the time when 
free neutrons are bound into nuclei, ${^3}$He and ${^7}$Li. Other than 
${^4}$He, the light element abundances in standard BBN reflect mainly the 
baryon fraction since the progress of the relevant reactions depends mainly 
on the concentration of particles. However, if we consider possible 
variations in fundamental parameters, different reactions and abundances may 
be affected in various ways, and in principle many different parameters 
could be bounded simultaneously \cite{MSW}.

Disadvantages of nucleosynthesis as a probe of fundamental parameters are the
large observational uncertainties; the complexity of the reaction network;
%, which is affected by gravitational, electromagnetic, weak and
%strong forces; 
and theoretical uncertainty in the dependence of nuclear reactions on QCD 
parameters. Advantages are the very large redshift (about $10^{10}$) making 
it the earliest direct test; 
%of any physical property; 
the independent WMAP estimate of the baryon fraction, which removes one 
unknown from the system; and the possibility of bounding many parameters at 
once, since nuclear reactions depend on the strong, electromagnetic and weak 
forces and freezeout is governed by the expansion of the Universe ({\em i.e.} 
gravity). Work is continuing with C.~Wetterich and S.~Stern to find a 
complete set of such bounds, incorporating recent advances in nuclear theory.

\paragraph{Theoretical issues} The only consistent way to introduce 
non-constant fundamental parameters in theory appears to be a cosmological
scalar field, which reduces the variation to a property of the particular
solution we inhabit, rather than the underlying theory. In order to have
non-negligible variation over cosmological times or distances, the scalar
should be extremely light, and either be ``rolling'' freely in a very shallow
potential, or be driven by the local matter density which is itself varying. 
Additionally there must be a coupling that induces a variation in observable
quantities. This leads to the possibility of long range forces due to the
scalar couplings, thus objects in free fall may not accelerate equally: this
violates the weak equivalence principle (WEP).

In the low energy limit the scalar couples in general to electromagnetic energy
and to the nucleon and electron masses: variations in $\alpha$ and $\mu$ are
then related to the size of couplings and the variation of the scalar. This 
variation is bounded above via the effect of its kinetic energy on the 
expansion of the Universe. Hence for a given variation in $\alpha$ or $\mu$
we expect the differential acceleration $\eta$ of two test bodies to be bounded
below.

We also require some relations between different scalar couplings to 
electromagnetism and matter: such relations arise from unified scenarios 
\cite{unification} where all observable variations arise from one underlying 
varying quantity such as a grand unified gauge coupling $\alpha_X$. Such 
relations can also be tested by comparing the sizes of $\Delta \ln \alpha$ 
and $\Delta \ln \mu$, if nonzero. We find for example \cite{me06}
\begin{equation}
\eta \geq \frac{ \Delta_{12} f_p}{2\dot{\bar{\phi}}_{\rm max}^2 } 
\frac{K}{c_2^2} \left( \frac{\dot{\mu}}{\mu} \right)^2\ \simeq
\frac{K}{c_2^2} \left( \frac{ \dot{\mu}/\mu }{ 3.7 \times 10^{-10} y^{-1} }
\right)^2
\end{equation}
where $\Delta_{12}f_p$ is the difference in proton fraction between two test
bodies, $\dot{\bar{\phi}}_{\rm max} \simeq 5\times 10^{-11}\,y^{-1}$ is the 
maximum allowed time derivative of the (dimensionless) scalar, and $K$ and 
$c_2^2$ are numbers of order 1 which depend on the details of the unified 
theory. If a nonzero variation does exist, the presence or absence of
WEP violation could test unified scenarios; conversely, bounds on WEP 
violation imply limits on the present rate of change of fundamental parameters,
comparable with those from atomic clocks.

\begin{theacknowledgments}
The author is supported by the {\it Impuls- und Vernetzungsfond der Helmholtz-Gesellschaft.}
\end{theacknowledgments}



%\bibliography{sample}

% \IfFileExists{\jobname.bbl}{}
%  \typeout{******************************************}
%  \typeout{** Please run "bibtex \jobname" to optain}
%  \typeout{** the bibliography and then re-run LaTeX}
%  \typeout{** twice to fix the references!}
%  \typeout{******************************************}

%\end{document}

%%%%%%%%%%%%%%%%%%%%%%%%%%%%%%%%%%%%%%%%%%%
%% The following lines show an example how to produce a bibliography
%% without the help of the BibTeX program. This could be used instead
%% of the above.
%%%%%%%%%%%%%%%%%%%%%%%%%%%%%%%%%%%%%%%%%%%

\end{document}